\documentclass[10pt]{bmc_article}

\usepackage{cite} 
\usepackage{url}  
\usepackage{ifthen}  
\usepackage{multicol}   
\usepackage[utf8]{inputenc} 
\usepackage{graphics}
\usepackage{graphicx}
\newboolean{publ}

\newenvironment{bmcformat}{\baselineskip20pt\sloppy\setboolean{publ}{false}}{\baselineskip20pt\sloppy}

\begin{document}
\begin{bmcformat}


\title{Integrating Prior Knowledge Into Prognostic Biomarker Discovery based on Network Structure}
 

\author{Yupeng Cun\email{Yupeng Cun - cun@bit.uni-bonn.de}
  , Holger Fr\"{o}hlich\correspondingauthor%
         \email{Holger Fr\"{o}hlich - frohlich@bit.uni-bonn.de}%
      }


\address{
    Bonn-Aachen International Center for Information Technology (B-IT), University of Bonn, Dahlmannstr. 2, 53113 Bonn, Germany  
}%

\maketitle


\begin{abstract}

\textbf{Background}: Predictive, stable and interpretable gene signatures are generally
seen as an important step towards a better personalized medicine.
During the last decade various methods have been proposed for that
purpose. However, one important obstacle for making gene signatures
a standard tool in clinics is the typical low reproducibility of
these signatures combined with the difficulty to achieve a clear
biological interpretation. For that purpose in the last years there
has been a growing interest in approaches that try to integrate
information from molecular interaction networks.

\textbf{Results}: We propose a novel algorithm, called FrSVM, which integrates
protein-protein interaction network information into gene selection
for prognostic biomarker discovery. Our method is a simple filter
based approach, which focuses on central genes with large
differences in their expression. Compared to several other competing
methods our algorithm reveals a significantly better prediction
performance and higher signature stability. Moreover, obtained gene
lists are highly enriched with known disease genes and drug targets.
We extendd our approach further by integrating information on
candidate disease genes and targets of disease associated Transcript Factors (TFs). 
\end{abstract}

\ifthenelse{\boolean{publ}}{\begin{multicols}{2}}{}



\section{Introduction}

During the last decade the topic ''personalized medicine`` has gained
much attention. One of the major goals is to identify reliable
molecular biomarkers that predict a patient's response to therapy,
including potential adverse effects, in order to avoid ineffective
treatment and to reduce drug side-effects and associated costs.
Prognostic or diagnostic biomarker signatures (mostly from gene
expression data, but more recently also from other data types, such
as miRNA, methylation patterns or copy number alterations) have been
derived in numerous publications for various disease entities. One
of the best known ones is a 70-gene signature for breast cancer
prognosis (mammaprint) by \cite{Veer2002}, which has gained FDA
approval.

A frequently taken approach to obtain a diagnostic or prognostic
gene signature is to put patients into distinct groups and then
constructing a classifier that can discriminative patients in the
training set and is able to predict well unseen patients. Well known
algorithms for this purpose are PAM \cite{Tibshirani2002}, SVM-RFE
\cite{Guyon2002}, Random Forests \cite{BreimanRandomForests2001}
or statistical tests, like SAM \cite{Tusher2001}, in combination
with conventional machine learning methods (e.g. Support Vector
Machines, k-NN, LDA, logistic regression, ...).

However, retrieved gene signatures are often not reproducible in the
sense that inclusion or exclusion of a few patients can lead to
quite different sets of selected genes. Moreover, these sets are
often difficult to interpret in a biological way
\cite{Goenen2009}. For that reason, more recently a number of
approaches have been proposed, which try to integrate knowledge on
canonical pathways, GO annotation or protein-protein interactions
into gene selection algorithms
\cite{Guo2005,Chuang2007,Rapaport2007,Yu2007,Lee2008,Taylor2009,Binder2009,Zhu2009,Johannes2010}.
A review on these and other methods can be found in
Cun and Fr\"{o}hlich\cite{Cun2012a}. The general hope is not only to make biomarker
signatures more stable, but also more interpretable in a biological
sense. This is seen as a key to making gene signatures a standard
tool in clinical diagnosis \cite{Blazadonakis2011}.

In this paper we propose a simple and effective filter based gene
selection mechanism, which employs the GeneRank algorithm
\cite{Morrison2005} to rank genes according to their centrality in
a protein-protein interaction (PPI) network and their (differential)
gene expression. It has been shown previously that deregulated
central genes have a strong association with the disease pathology
in cancer \cite{Wang2011graphCentrality}. Our method uses the span
rule \cite{Chapelle2002} as a bound on the leave-one-out error of
Support Vector Machines (SVMs) to filter the top ranked genes and
construct a classifier. It is thus conceptually and computationally
much simpler than our previously proposed RRFE algorithm
\cite{Johannes2010}, which used a reweighting strategy of the SVM
decision hyperplane. We here demonstrate that our novel method,
called FrSVM, not only significantly outperforms RRFE, PAM, network
based SVMs \cite{Zhu2009}, pathway activity classification
\cite{Lee2008} and average pathway expression\cite{Yu2007}, but
that it also yields extremely reproducible gene signatures.

In a second step we investigate, in how far our approach can be
improved further by incorporating potential disease genes or targets
of transcription factors, which were previously found to be enriched
in known disease genes. It turns out that the combination with
candidate disease genes can further improve the association to
biological knowledge.

\section{Methods}
\subsection{Datasets}

We retrieved two breast cancer \cite{Ivshina2006,Schmidt2008} and
one prostate cancer \cite{Sun2009} dataset from the NCBI GEO data
repository \cite{Barrett2011}. Moreover, TCGA \cite{TCGA2011oc}
was used to obtain an additional dataset for ovarian cancer
(normalized level 3 data). All data were measured on Affymetrix
HGU133 microarrays (22,283 probesets). Normalization was carried via
FARMS (breast cancer datasets - \cite{Hochreiter2006}) and
quantile normalization (prostate cancer dataset -
\cite{Bolstad2003}), respectively. As clinical end points we
considered metastasis free (breast cancer) and relapse free (ovarian
cancer) survival time after initial clinical treatment. For ovarian
cancer only tumors with stages IIA - IV and grades G2 and G3 were
considered, which after resection revealed at most 10mm residual
cancer tissue and responded completely to initial chemo therapy.

Survival time information was dichotomized into two classes
according whether or not patients suffered from a reported relapse /
metastasis event within 5 (breast) and 1 year (ovarian),
respectively. Patients with a survival time shorter than 5 / 1
year(s) without any reported event were not considered and removed
from our datasets. For prostate cancer we employed the class
information provided by \cite{Sun2009}. A summary of our datasets
can be found in Table 1.

\subsection{Protein-Protein Interaction (PPI) Network}

A protein interaction network was compiled from a merger of all
non-metabolic KEGG pathways \cite{Kanehisa2008}-  only gene-gene
interactions were considered -- together with the Pathway Commons
database \cite{Cerami2011}, which was downloaded in tab-delimited
format (May 2010). The purpose was to obtain an as much as possible
comprehensive network of known protein interactions. For the Pathway
Commons database the $SIF$ interactions INTERACTS\_WITH and
STATE\_CHANGE were taken into
account\footnote[1]{http://www.pathwaycommons.org/pc/sif\_interaction\_rules.do}
and any self loops removed. For retrieval and merger of KEGG
pathways, we employed the R-package KEGGgraph \cite{Zhang2009}. In
the resulting network  graph (13,840 nodes with 397,454 edges) we
had directed as well as undirected edges. For example, a directed
edge $A\to B$ could indicate that protein $A$ modifies protein $B$
(e.g. via phosphorylation). An undirected edge $A-B$ implies a not
further  specified type of direct interaction between $A$ and $B$.
Nodes in this network were identified via Entrez gene IDs.

The R package, \textit{hgu133a.db} \cite{Carlson2009}, was
employed to map probe sets on the microarray to nodes in the
PPI-network. This resulted in a protein-protein interaction network
matrix of dimension $8876\times8876$, because several probe sets can
map to the same  protein in the PPI-network. Accordingly, expression
values for probesets on the microarray that mapped to the same gene
in the network were averaged. Probesets, which could not be mapped
to the PPI network, were ignored for all network based approaches
except for RRFE, which according  to Johannes et al \cite{Johannes2010}, assigns a minimal gene rank to them.

\subsection{Gene Selection with PPI Information (FrSVM)}

The GeneRank algorithm described in Morrison et al \cite{Morrison2005} is an
adaption of Google's PageRank algorithm. It combines gene expression
and protein-protein interaction information to obtain a ranking of
genes by solving the linear equation system
\begin{equation}
\left(\mathbf{I}-d\mathbf{W}\mathbf{D}^{-1}\right)\mathbf{r}=(1-d)\mathbf{e}\label{eq:GeneRank}
\end{equation}
where $\mathbf{W}$ denotes the adjacency matrix of the PPI network,
$\mathbf{D}$ is a diagonal matrix consisting of the node degrees and
$d$ a damping factor weighting (differential) gene expression
$\mathbf{e}$ against network information. As suggested in
Morrison et al \cite{Morrison2005} we set $d=0.85$ here. The general idea of the
algorithm is to give preference to proteins, which are central in
the network (similar to web pages with many links) on one hand and
have a high difference in their expression on the other hand.

As a score for differential gene expression (vector $\mathbf{e}$) we
employed the absolute value of t-statistics here. That means we
conducted for each probeset a t-test and then looked at the absolute
t-value to assign weights to nodes in the PPI network. This in turn
allowed us to apply GeneRank to calculate a rank for each probeset.
We then filtered the top ranked 10, 11, ..., 30\% of all probesets
mapping to our PPI network and each time trained a Support Vector
Machine (SVM). We used the span rule \cite{Chapelle2002} to
estimate an upper bound on the leave-one-out error in a
computationally efficient way. This was only done on the training
data and allowed us to select the best cutoff value for our filter.
At the same time we could use the span rule also to tune the soft
margin parameter $C$ of the SVM in the range
$10^{-3},10^{-2},...,10^{3}$. Out approach is called \textbf{FrSVM}
in the following.

\subsection{Using Candidate Disease Genes}

For many diseases several associated genes are known. Based on this
information it is possible to prioritize candidate genes via their
similarity to known disease genes: Schlicker et al \cite{Schlicker2010} proposed a
mechanism to compute similarities of gene products to candidate
genes based on their Gene Ontology (GO) information. The Endeavour
software \cite{Endeavor2008} employs a different algorithm to rank
candidate genes based on their proximity in annotation space by
combining information sources like GO, KEGG, text and others.

We here tested a combination of the propose FrSVM algorithm with
both disease gene prioritization approaches (Endeavour and GO
similarity): We selected the top ranked p\% genes according to FrSVM
as well as according to Endeavour and GO Similarity. The union of
both sets was then used for SVM training. For Endeavour we
considered GO, KEGG, text and sequence motifs as information
sources. Information on disease related genes was obtained from the
DO-light ontology \cite{Osborne2009}. GO functional similarity was
computed via the method proposed in \cite{Mistry2008} using the
web tool FunSimMat \cite{Schlicker2008}, which uses the NCBI OMIM
database for disease gene annotation. The combination of FrSVM with
Endeavour is called \textbf{FrSVM\_EN}, and the combination with
functional GO similarities is called \textbf{FrSVM\_FunSim}
accordingly.

In addition to FrSVM\_EN and FrSVM\_FunSim we also considered to use
the top ranked candidate disease genes only (without any further
network information). The corresponding approaches are principally
equivalent to FrSVM from the methodological point of view (just
another ranking is used) and are called \textbf{EN} and
\textbf{FunSim}, respectively.

\subsection{Using Targets of Enriched Transcription Factors}

A major factor influencing gene expression are transcription factors
(TFs). We performed a hypergeometric test looked for enriched TF
targets in disease associated genes (FDR cutoff 5\%). Only probesets
mapping to targets of enriched TFs were then taken into account to
conduct a subsequent FrSVM training. We refer to this method as
\textbf{FrSVM\_TF}. Again, information on disease relation of genes
was obtained from the DO-light ontology. A TF-target gene network
was compiled by computing TF binding affinities to promoter
sequences of all human genes according to the TRAP model
\cite{Roider2007} via the author's R implementation. Upstream
sequences of genes were retrieved here from the ENSEMBL database via
biomaRt \cite{Guberman2011}. We assumed that promoter sequences
were located in the range 0 - 2Kbp upstream to the transcription
start site of a gene. As trustworthy TF targets we considered those,
for which a Holm corrected affinity p-value smaller than 0.01 was
reported. In conclusion we found 6334, 8196 and 5866 probesets
(having enriched binding sites of 33, 35 and 24 TFs) for breast,
prostate and ovarian cancer.

\subsection{Classification Performance, Signature Stability and Biological Interpretability}

In order to assess the prediction performance we performed a 10
times repeated 10-fold cross-validation on each dataset. That means
the whole data was randomly split into 10 fold, and each fold
sequentially left out once for testing, while the rest of the data
was used for training and optimizing the classifier (including gene
selection, hyper-parameter tuning, standardization of expression
values for each gene to mean 0 and standard deviation 1, etc.). The
whole process was repeated 10 times. It should be noted extra that
also standardization of gene expression data was only done on each
training set separately and the corresponding scaling parameters
then applied to the test data. The area under receiver operator
characteristic curve (AUC) was used here to measure the prediction
accuracy, and the AUC was  calculated by R-package ROCR
\cite{Sing2005}. To assess the stability of features selection
methods, we computed the selecticomparableon frequency of each gene within the
10 times repeated 10-fold cross-validation procedure. In an ideal case
probsets would be selected consistently, i.e. all probeset chosen
100 times. The more the probeset selection profile (which is essentially
a histogram) resembles this ideal case the better. In order to capture
this behavior numerically we defined a so-called \emph{stability index}
(SI) defined as\[
SI=\sum_{i\in\{10,20,...,100\}}i\cdot f(i)\]
where $f(i)$ denotes the fraction of probsets that have been selected$>i-10$
and $\leq i$ times. Please note that $\sum_{i}f(i)=1$. $SI$ represents
a weighted histogram count of selection frequencies. Obviously, the
larger $SI$ the more stable the algorithm is. In the optimal case
$SI=100$.

We also looked, in how far signatures obtained by training the
classifier on the whole dataset could be related to existing
biological knowledge. For this purpose we looked for enriched
disease related genes and known targets of therapeutic compounds via
a hypergeometric test. For disease related genes we made use of the
tool ``FunDO'' \cite{Osborne2009}. Multiple testing correction is
done here via Bonferroni's method. The list of therapeutic compounds
and their known targets was retrieved via the software
MetaCore\texttrademark (GeneGo Inc.) and is available in the
supplements.

\section{Results and Discussion}

\subsection{FrSVM improves Classification Performance and Signature Stability }

We compared the prediction performance of our proposed FrSVM method
to PAM \cite{Tibshirani2002}, average gene expression of KEGG
pathways (\textbf{aveExpPath}, \cite{Yu2007}), pathway activity
classification (\textbf{PAC}, \cite{Lee2008}), network-based SVM
(\textbf{networkSVM}, \cite{Zhu2009}) and reweighted recursive
feature elimination (\textbf{RRFE}, \cite{Johannes2010}). For
aveExpPat we first conducted a global test \cite{goeman04} to
select pathways being significantly associated with the phenotype
(FDR cutoff 1\%) and then computed the mean expression of genes in
these pathways.

Initially we only used PPI information for our FrSVM approach and
found a clear improvement of AUC values for FrSVM compared to all
other tested methods (Figure 1). This
visual impression was confirmed via a two-way ANOVA analysis (using
method, dataset as well as their interaction term as factors) with
Tukey's post-hoc test, which revealed a significantly increased AUC
for FrSVM with p $<$ 1e-6 in all cases.

We further inspected the frequencies, by which individual probesets
were selected by each of the tested methods (Figure 2) as well as the stability indices (Figure 2b). This analysis showed that FrSVM
selected probesets in a very stable manner (only comparable to
networkSVM). The fraction of consistently selected probesets ranged
from \textasciitilde{}40\% (ovarian cancer) to \textasciitilde{}70\%
(Schmidt et al. breast cancer dataset). Interestingly these
consistently selected genes typically showed a highly significant
differential expression, which was assessed via SAM
\cite{Tusher2001} here. For example, 60\% of all consistently
selected probesets in the Schmidt et al. dataset had a q-value $<$
5\%. This illustrates the behavior of FrSVM to focus on genes with
large differences in their expression between the two compared
groups, which are central in the PPI network.

\subsection{Clear Association to Biological Knowledge}

We trained each of our test methods on complete datasets to retrieve
final signatures, which we tested subsequently for the enrichment of
disease related genes and known drug targets (Figure 3 and Figure 4). This analysis
showed that FrSVM derived signatures can be clearly associated to
biological knowledge. The degree of enrichment was only comparable
with aveExpPath and RRFE, which have previously been found to yield
clearly interpretable signatures \cite{Cun2012}.

\subsection{PPI Network Integration Helps Most}

We went on to test, how much the performance of FrSVM would be
affected by integrating candidate disease genes or restricting
selectable probesets to targets of enriched TFs. Generally,
incorporation of network knowledge appeared to yield a better
prediction performance than only using candidate disease genes
Figure 5, $p < 0.01$, two-way ANOVA
with Tukey's post-hoc test). No significant benefit of additionally
integrating candidate disease genes or targets of enriched TFs into
FrSVM could be observed in terms of AUC values or signature
stabilities Figure 5b). However, FrSVM\_EN showed a clearer association to
disease genes than FrSVM (Figure 6). This is not surprising, because the method explicitely integrates the top ranked candidate disease genes.

%
%

\section{Conclusion}

We proposed a simple and effective filter based algorithm to
integrate PPI network information into prognostic or diagnostic
biomarker discovery based on a modification of Google's PageRank
algorithm. The method favors genes, which on one hand show a large
difference in their expression (high absolute t-score) and on the
other hand are central in the network. It has been shown previously
that such genes are often associated to the disease phenotype
\cite{Wang2011graphCentrality}. Our approach significantly
outperformed several other classification algorithms in terms of
prediction performance and signature stability on four datasets.
Moreover, it yielded signatures showing a very clear relation to
existing biological knowledge. Additional integration of potential
disease genes could further enhance this association, but
nonetheless did not improve prediction performance or signature
stability. PPI network integration appeared to be more effective
than integration of candidate disease genes. Using only targets of
TFs, which were previously found to be enriched in known disease
genes, did not reveal any significant improvement. However, from a
computational point of view this approach might still be
interesting, because the set of candidate probesets is significantly
restricted before any time consuming machine learning algorithm is
applied.

In conclusion, our method offers a computationally cheap and
effective mechanism to include prior knowledge into gene selection
for biomarker discovery.

\section*{Acknowledgement}
This work was partially supported by the state of NRW via the B-IT
research school. We would like to thank Khalid Abnaof for providing 
the data of TFs genes. 



\newpage
{\ifthenelse{\boolean{publ}}{\footnotesize}{\small}
 \bibliographystyle{bmc_article}  
  \bibliography{frReferences} }     


\ifthenelse{\boolean{publ}}{\end{multicols}}{}


\newpage
\section*{Figures}
  \subsection*{Figure 1 - Prediction performance of FrSVM in comparison to other methods in
terms of area under ROC curve (AUC).}
  \includegraphics{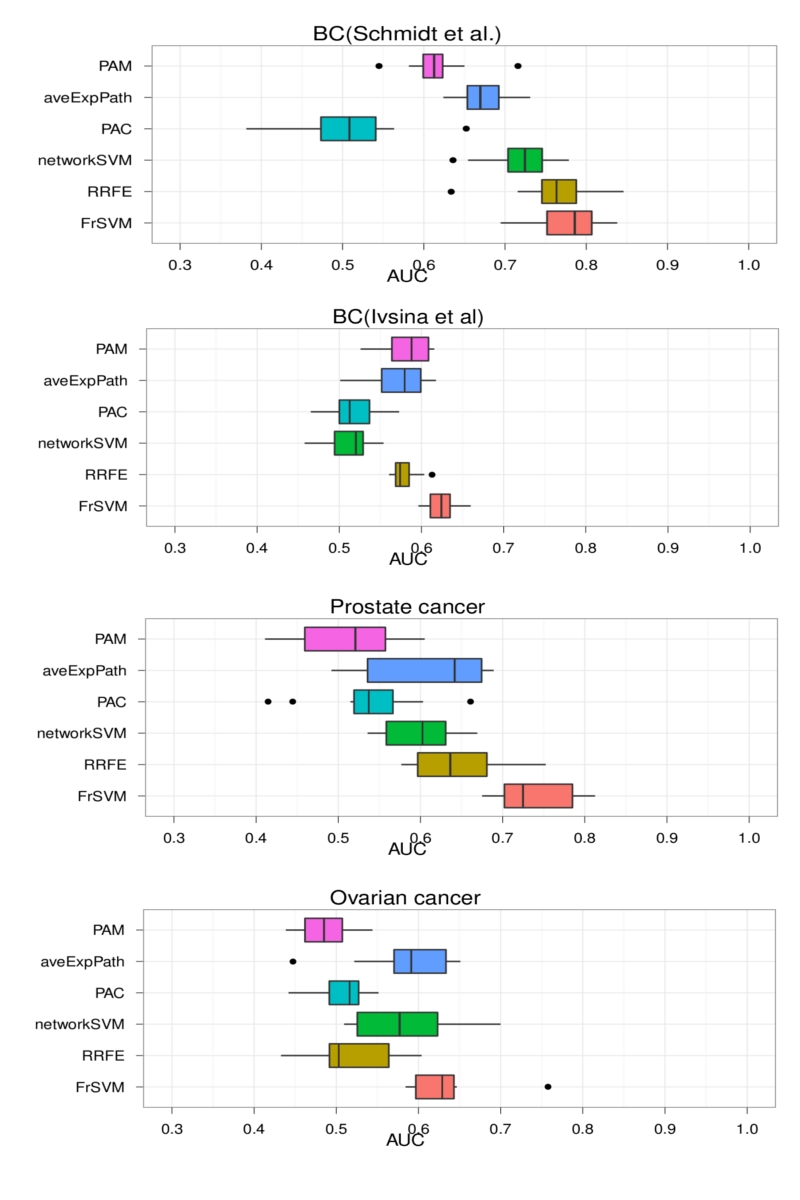}
\newpage

  \subsection*{Figure 2 - Fraction of probesets that were selected 1 - 10, 11 - 20, ..., 99
- 100 times within the 10 times repeated 10-fold CV procedure.} 
\includegraphics[scale=0.5]{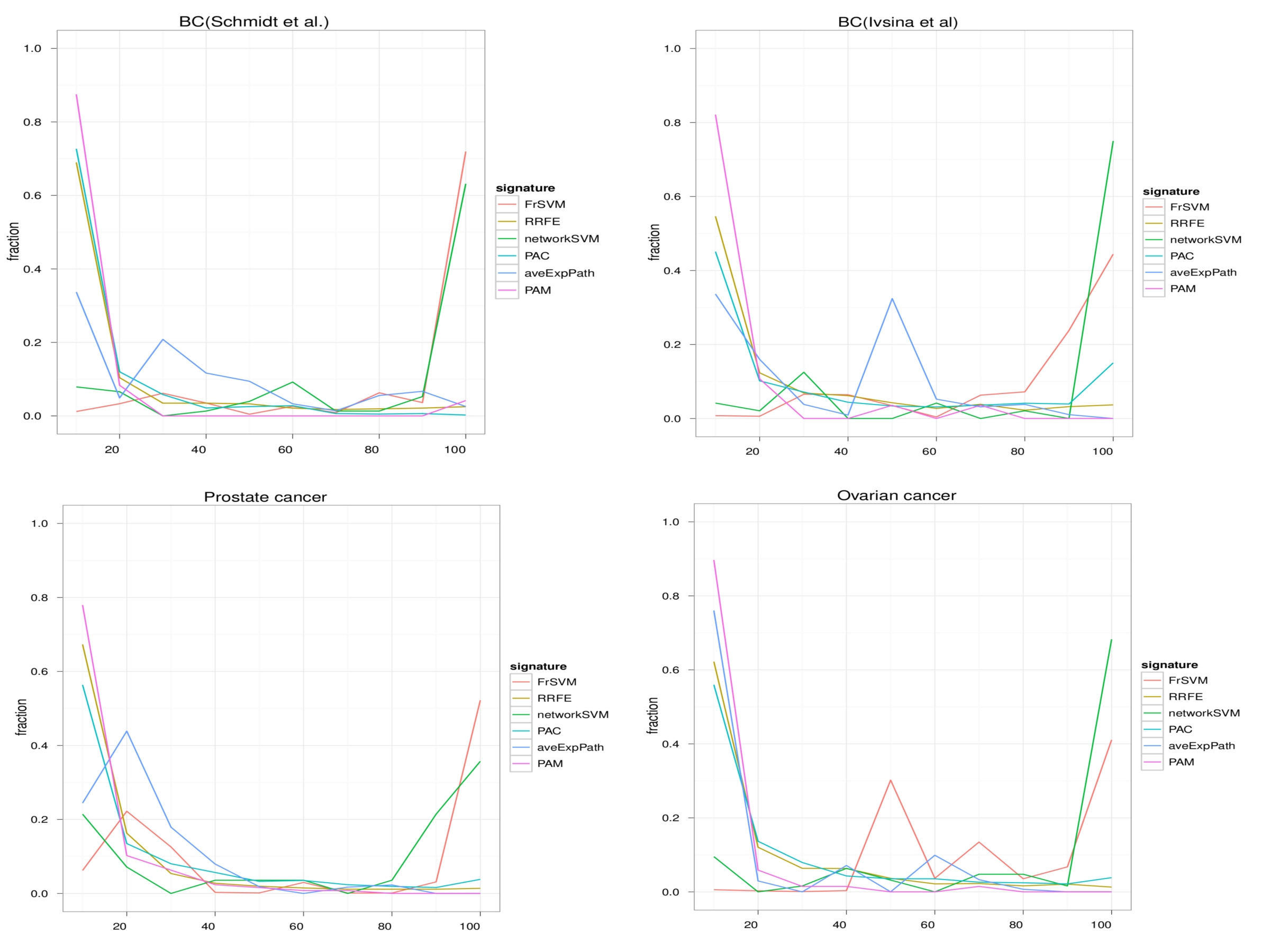}

\newpage
 \subsection*{Figure 3 - Stability indices (SI) of compared methods.}
\includegraphics{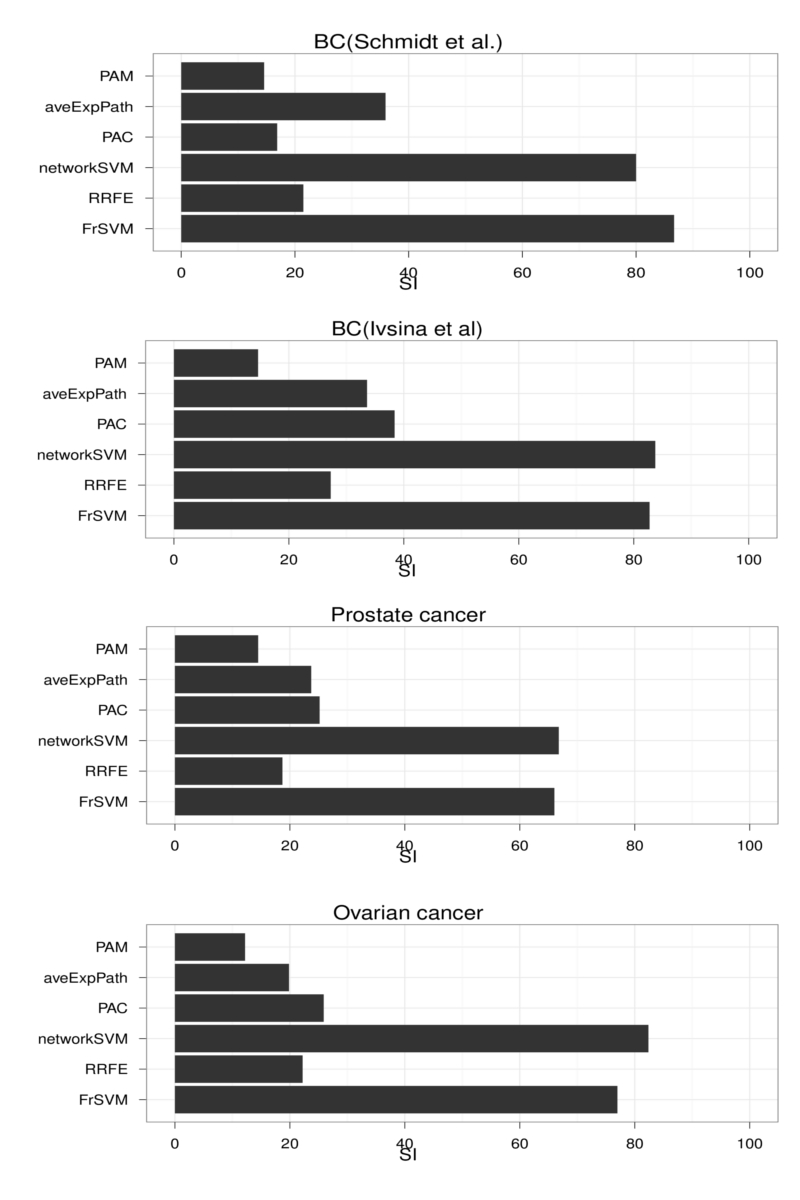}

\newpage
 \subsection*{Figure 4 - Enrichment of signatures with disease related genes.}
\includegraphics{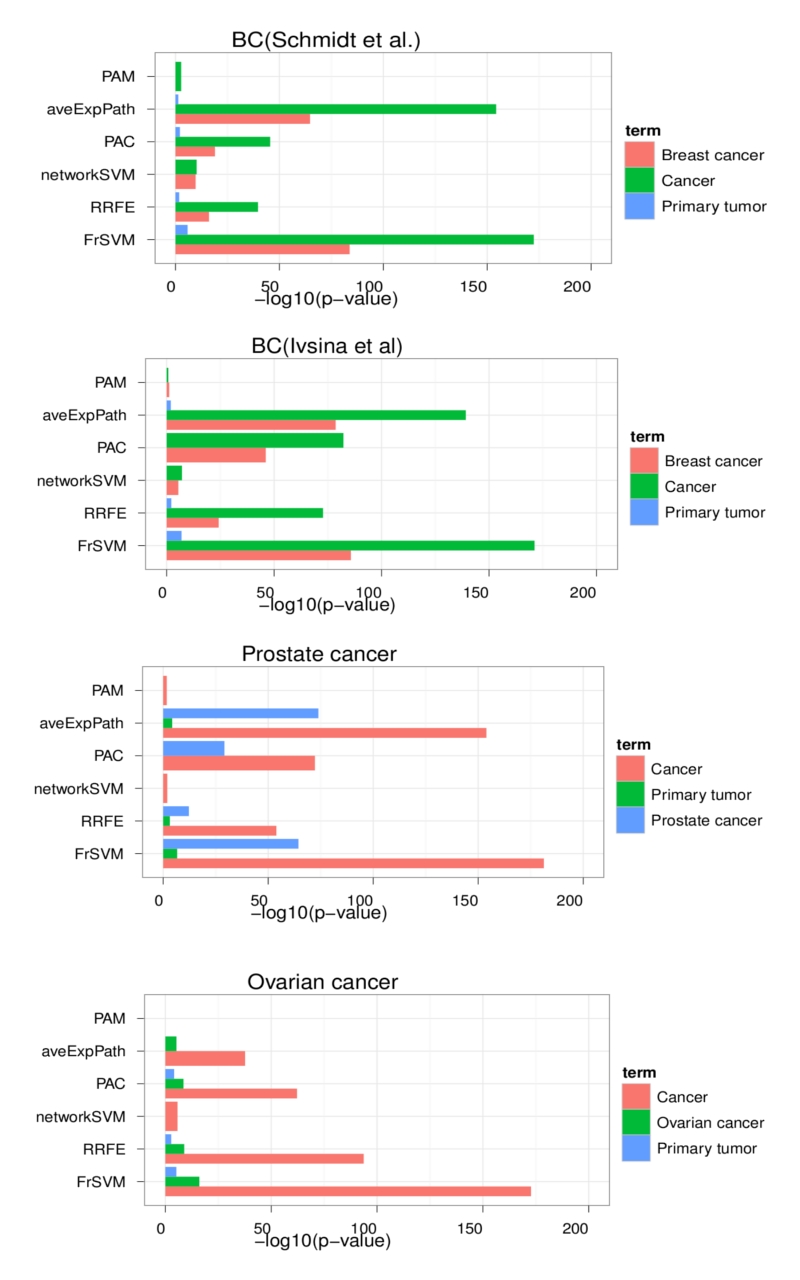}

\newpage
 \subsection*{Figure 5 - Enrichment of signatures with known drug targets.}
\includegraphics{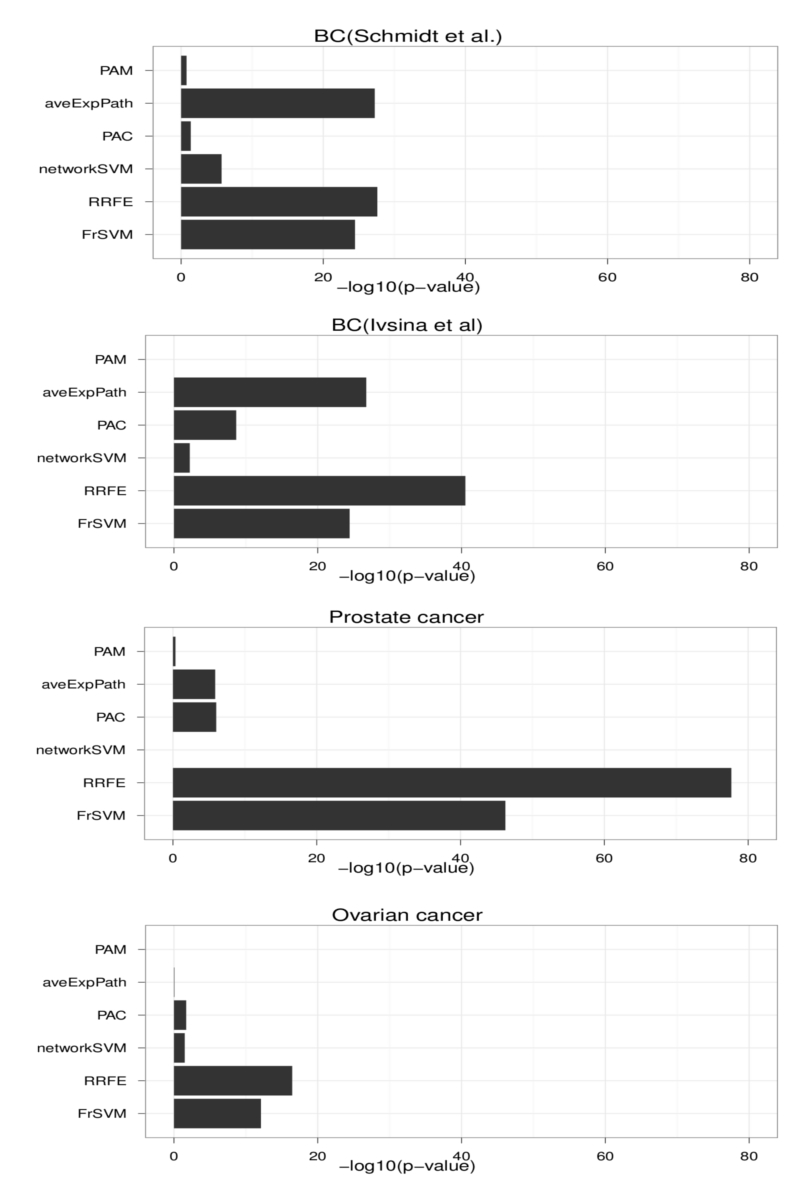}

\newpage 
\subsection*{Figure 6 -  Effect of integrating prior information in addition to protein interac-
tions into FrSVM: prediction performance.}
\includegraphics{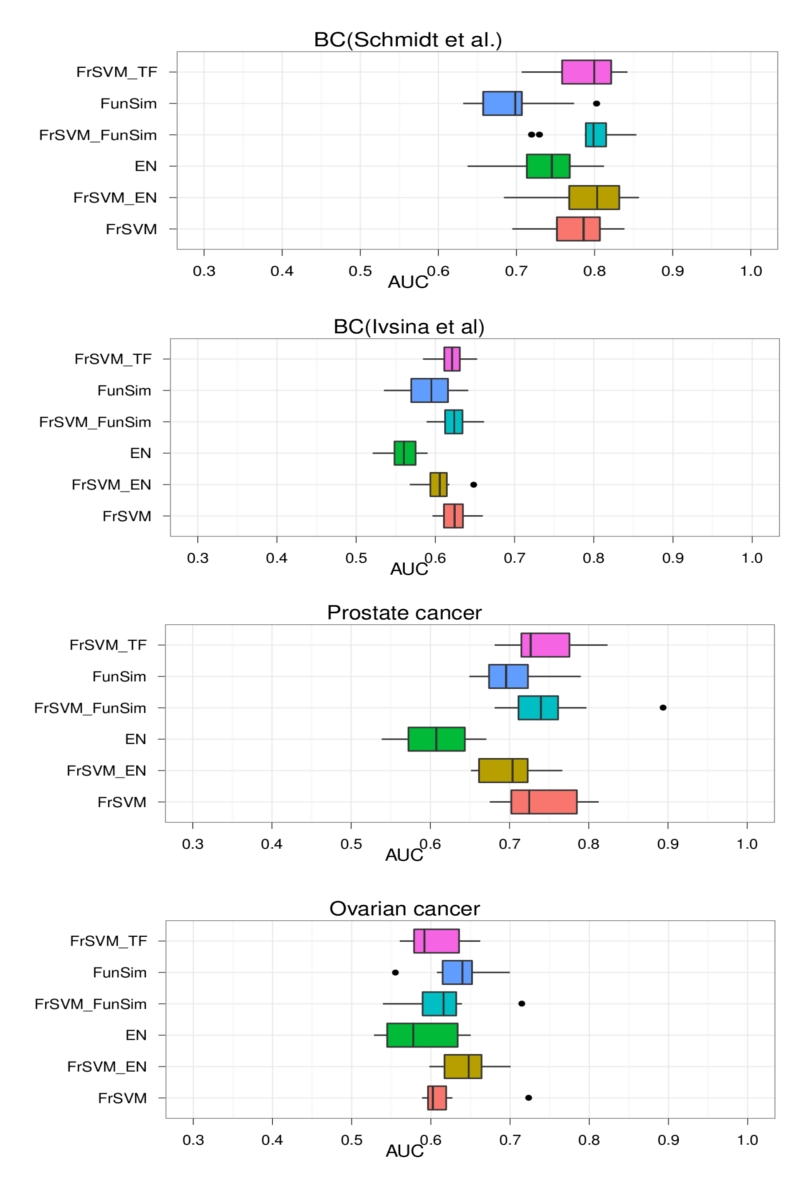}

\newpage
 \subsection*{Figure 7 - Effect of integrating prior information in addition to protein interac-tions into FrSVM: stability index}
\includegraphics{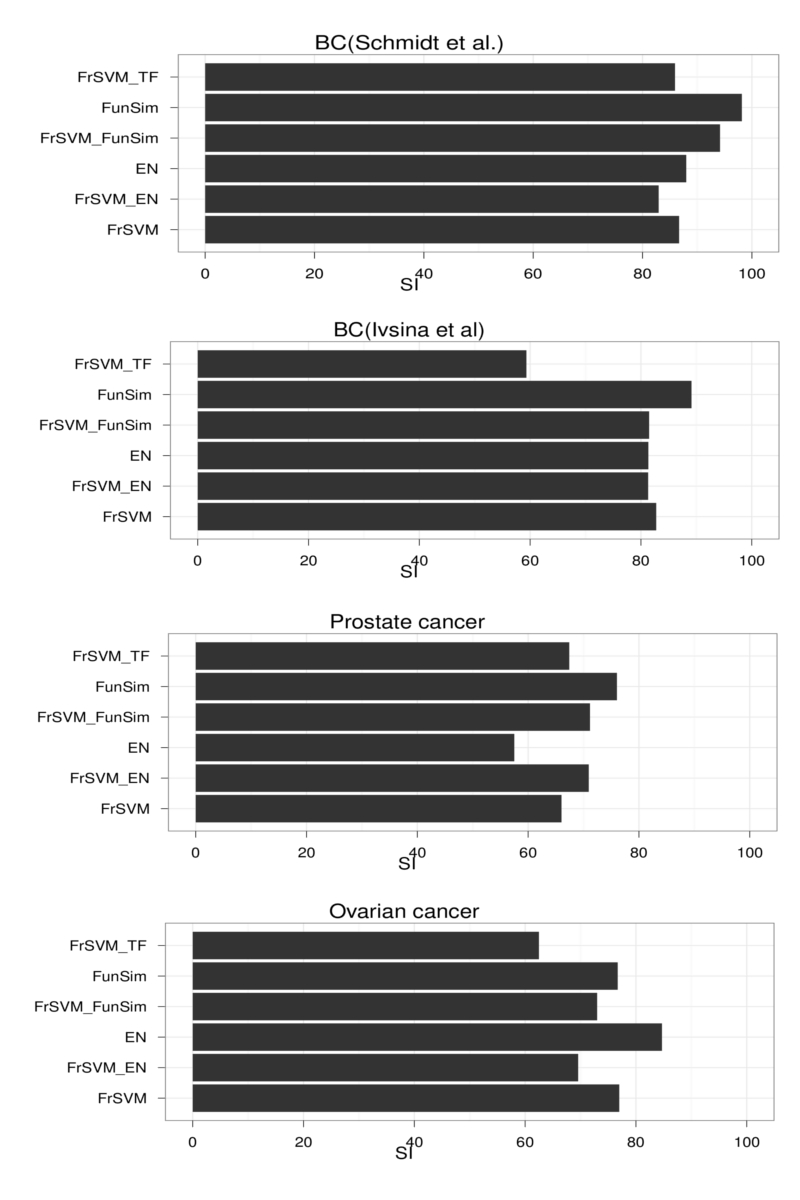}

\newpage
 \subsection*{Figure 8 - Enrichment of signatures with disease related genes after integration
of prior information additional to protein interactions.} 
\begin{figure}[htb]
\includegraphics{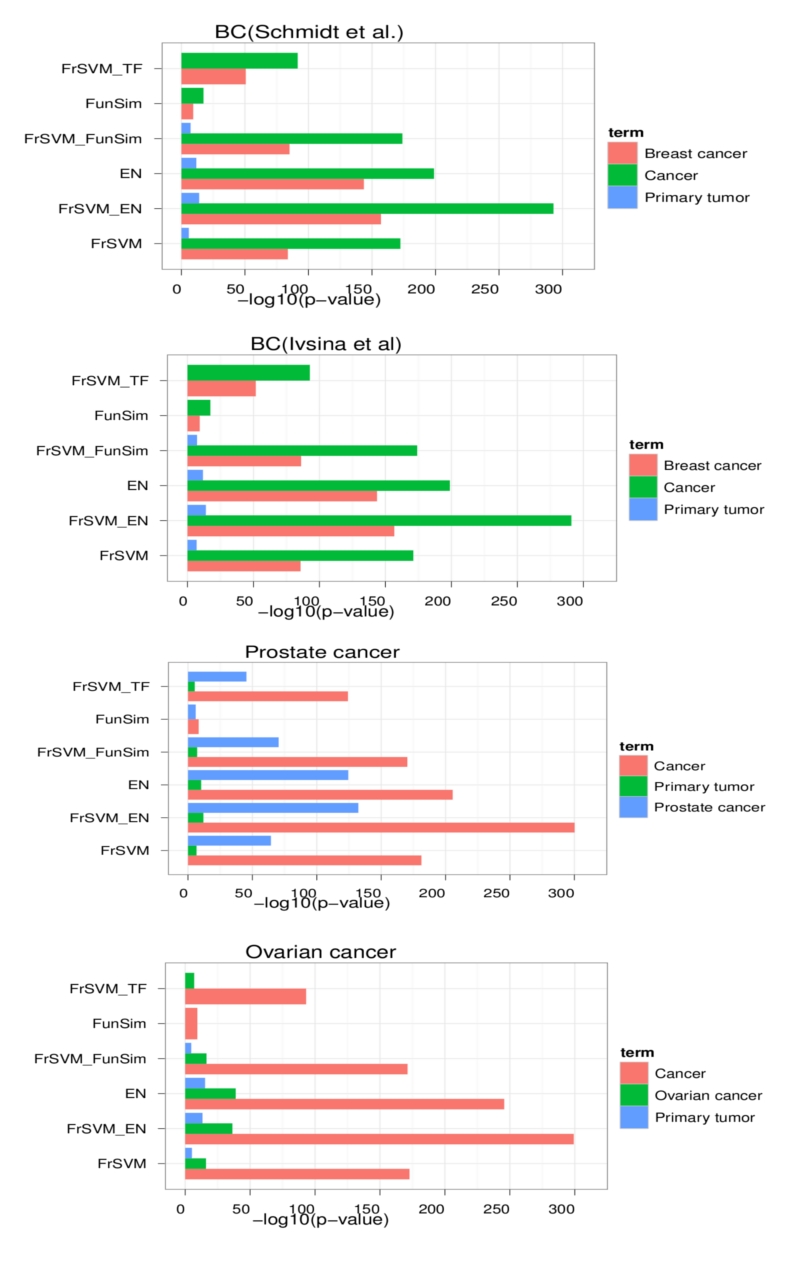}
\end{figure}




\newpage
\section*{Tables 1 - Overview about employed datasets} 
\begin{table}
\begin{tabular}{|c|c|c|c|c|c|}
\hline 
GEOid & examples & cancer type & Predict label & Positive & Data source\tabularnewline
\hline
\hline 
GSE11121 & 182  & Breast Cancer  & DFS smaller than 5 years V.S. 5 years  & 28 & Schmidt et al \cite{Schmidt2008}\tabularnewline
\hline 
GSE4922  & 228 & Breast cancer & DFS smaller than 5 years V.S. 5 years & 69 & Ivshina et al \cite{Ivshina2006}\tabularnewline
\hline 
TCGA  & 135 & Ovarian Cancer  &  relapse free survival $>$ 1y & 35  & TCGA \cite{TCGA2011oc}\tabularnewline
\hline 
GSE25136  & 79 & Prostate Cancer  & Recurrent V.S. Non-Recurrent & 40 & Sun et al \cite{Sun2009}\tabularnewline
\hline
\end{tabular}
\end{table}


\end{bmcformat}
\end{document}